\documentclass[a4paper,11pt]{article}
\usepackage{latexsym}
\usepackage{amsmath}
\usepackage{amssymb}
\usepackage{graphicx}
\usepackage{wrapfig}
\usepackage{fancybox}
\usepackage{bm}
\title{The separation of the chiral and deconfinement phase transitions in the curved space-time}
\author{S.Sasagawa and H.Tanaka \\\\Department of Physics, Rikkyo University, Tokyo 171-8501, Japan}
\date{}
\usepackage{fancyhdr}
\makeatletter
\def\firstpage{\hfill RUP-12-9}
\def\ps@titlepage{%
   \@oddhead{\hfil\firstpage\hfil}%
   \let\@evenhead\@oddhead
   \def\@oddfoot{\hfil}%
   \let\@evenfoot\@oddfoot
   \let\@mkboth\@gobbletwo}
\makeatother
\begin{document}
\vspace{1em}
\maketitle

\thispagestyle{titlepage}

\vspace{1em}
\begin{abstract}

\noindent We calculated the chiral condensate and the dressed Polyakov loop in the space-time $R\times S^{3}$ and $R\times H^{3}$. The chiral condensate is the order parameter for the chiral phase transition, and the dressed Polyakov loop is the order parameter for the deconfinement phase transition. When there is a current mass, critical points for the chiral and deconfinement phase transitions are different in the crossover region. We show that the difference is changed by the gravitational effect.

\end{abstract}

\vspace{1em}
$\\[0.28cm]$

\vspace{1em}
\section{Introduction}\label{sec:njl}

Quantum chromodynamics (QCD) in non-perturturbative region has two important properties. One is breaking of chiral symmetry, another is confinement. It is believed that chiral symmetry breaking is restored and the confinement phase transforms into the deconfinement phase at finite temperature and density. The underlying QCD phase transition at temperature and density is constructed by these. The QCD phase transition is searched at GSI, CERN, SPS, RHIC, and LHC (see, eg., Ref.\ \cite{rf:ex}). However, we understand little about the explicit relation between the chiral and deconfinement transitions. For example, we do not have the satisfactory effective model that can describe the chiral and deconfinement phase transitions. In addition, order parameters for deconfinement in the functional method are not simpler than one for chiral symmetry breaking \cite{rf:de}.

As a theoretical approach to understand the phase structure of QCD, the lattice QCD simulation is a powerful method. However, there are different results for the chiral and deconfinement phase transitions. The Wuppertal-Budapest group showed three critical temperature for the case of $N_{f}=2+1$. For the chiral phase transition of $u$ and $d$ quarks, the critical temperature is $T_{c}^{\chi(u,d)}=151$ MeV. For $s$ quark, critical temperature is $T_{c}^{\chi(s)}=175$ MeV. For the deconfinement phase transition, the critical temperature is $T_{c}^{d}=176$ MeV \cite{rf:Wu}. Thus, two phase transitions have different critical temperature. On the other hand, Bielefeld group showed the result that the critical temperature of two phase transitions coincide at $T_{c}=154$ MeV \cite{rf:Bie}. In this way, the lattice simulation still can not explain the association between the chiral and deconfinement phase transitions \cite{rf:Wu2}.

In this paper, we use the Nambu-Jona-Lasinio (NJL) model to investigate the chiral and deconfinement phase transitions. The NJL model can describe chiral symmetry properties on QCD \cite{rf:hatsu,rf:Buba}. However, since the NJL model lacks confinement, the deconfinement phase transition can not be described. On the other hand, the Polyakov loop extended NJL (PNJL) model can describe the chiral and deconfinement phase transitions \cite{rf:pnjl1,rf:pnjl2,rf:pnjl3,rf:pnjl4,rf:pnjl5,rf:pnjl6}. Although the PNJL model is one way to study the deconfinement phase transition, we do not use the PNJL model. Thus, we use the NJL model in the absence of any confinement mechanism. Instead, we calculate the dressed Polyakov loop \cite{rf:dress}. Since the dressed Polyakov loop has center symmetry, this is used as the order parameter for the deconfinement phase transition. As shown in Ref.\ \cite{rf:njldr}, the dressed Polyakov loop without any confinement mechanism behaviors like an order parameter. (The PNJL model with the dressed Polyakov loop is shown in Ref\ \cite{rf:pnjldr}.) In this method, the critical points for the deconfinement phase transition are coincident with the chiral phase transition in the chiral limit $m=0$. By contrast, in the case of $m\neq 0$, the critical points are not coincident in the crossover region \cite{rf:njldr}. This shows that there exists an intermediate region. (If the chiral and deconfinement phase transitions are of first order, two phase transitions should appear simultaneously.) An investigation by the Schwinger-Dyson equation at finite temperature and density also shows that there is the difference between the chiral and deconfinement phase transitions \cite{rf:sde}.

To investigate an association between the chiral and deconfinement phase transitions from a different standpoint, we add a gravitational effect. For example, the separation of the chiral and deconfinement phase transitions in strong magnetic field is shown in Ref.\ \cite{rf:njlmag}. Similarly, we investigate whether the separation of the chiral and deconfinement phase transitions are caused by a gravitational effect. The NJL model in curved space-time was investigated in recent year \cite{rf:njlg00,rf:njlg01,rf:njlg02,rf:njlg0,rf:njlg1,rf:njlg11,rf:njlg2,rf:njlg3}, and it was found that chiral symmetry breaking is restored. Thus, the chiral phase transition is affected by the curvature. Similarly, the deconfinement phase transition also may be affected. To investigate a gravitational effect for the chiral and deconfinement phase transitions, we try two cases: the positive curvature space-time $R\times S^{3}$ (the Einstein universe) \cite{rf:njlg2}, and the negative curvature space-time $R\times H^{3}$ (the hyperbolic space) \cite{rf:njlg3}. To calculate the thermodynamic potential in these cases, we use the mean field approximation.

This paper is organized as follows. In section\ \ref{sec:njl},  we review the NJL model in curved space-time. In section\ \ref{sec:phi}, we introduce the dressed Polyakov loop, and calculate the $\phi$ dependent thermodynamic potentials in the space time $R\times S^{3}$ and $R\times H^{3}$. We show numerical results in section\ \ref{sec:nu}, and a summary is found in section\ \ref{sec:sum}.

\vspace{1em}
\vspace{1em}
\section{NJL model in curved space-time}\label{sec:njl}

The $SU(N_{c})$ NJL Lagrangian with $u$ and $d$ quarks (the number of flavors $N_{f}=2$) takes the form,

\begin{equation}
\displaystyle \mathcal{L}=\overline{q}(i\gamma_{\mu}\partial^{\mu}-m+\mu\gamma_{0})q+\frac{G}{2N_{c}}[(\overline{q}q)^{2}+(\overline{q}i\gamma^{5}\bm{\tau} q)^{2}],
\end{equation}

\noindent where $N_{c}=3$ is the number of colors, $\bm{\tau}=(\tau_{1},\tau_{2},\tau_{3})$ are Pauli matrices in the flavor space, and $\mu$ is the quark chemical potential. We do not introduce the isospin chemical potential.

In 4-dimensional curved space, the gamma matrices are given by the following relations \cite{rf:spinor}

\begin{equation}
\{\gamma_{\mu},\gamma_{\nu}\}=2g_{\mu\nu}\ ,\ \{\gamma_{a},\gamma_{b}\}=2\eta_{ab},
\end{equation}

\noindent with the line element

\begin{equation}
ds^{2}=g_{\mu\nu}dx^{\mu}dx^{\nu}=\eta_{ab}e_{\mu}^{a}e_{\nu}^{b}dx^{\mu}dx^{\nu}.
\end{equation}

\noindent$e_{\mu}^{a}$ is the tetrad and $\eta_{ab}$ is the Minkowski metric, diag$(1,-1,-1,-1)$. Using the spin connection $\omega_{\mu}^{ab}$, the spinor covariant derivative $\nabla_{\mu}$ is expressed as

\begin{equation}
\displaystyle \nabla_{\mu}=\partial_{\mu}+\frac{1}{2}\omega_{\mu}^{ab}\sigma_{ab},\ \sigma_{ab}=\frac{1}{4}[\gamma_{a},\gamma_{b}].
\end{equation}

\noindent By these relations, the action in curved space-time is given by

\begin{equation}
S=\displaystyle \int d^{4}x\sqrt{-g}(\overline{q}(i\gamma_{\mu}\nabla^{\mu}-m+\mu\gamma_{0})q+\frac{G}{2N_{c}}[(\overline{q}q)^{2}+(\overline{q}i\gamma^{5}q)^{2}]),\label{eq:1}
\end{equation}

\noindent where $g=\det g_{\mu\nu}$. By applying the bosonization procedure, the action (\ref{eq:1}) is rewritten as

\begin{equation}
S=\displaystyle \int d^{4}x\sqrt{-g}(\overline{q}(i\gamma_{\mu}\nabla^{\mu}-m-\sigma-i\gamma_{5}(\bm{\tau}\cdot\bm{\pi})+\mu\gamma_{0})q-\frac{3}{2G}(\sigma^{2}+\bm{\pi}^{2})).
\end{equation}

\noindent Boson fields $\sigma$ and $\bm{\pi}$ satisfy the equations of motion,

\begin{equation}
\displaystyle \sigma=-\frac{G}{3}\overline{q}q,\ \bm{\pi}=-\frac{G}{3}\overline{q}i\gamma_{5}\bm{\tau} q.
\end{equation}

We use the mean field approximation, $\sigma=\langle\sigma\rangle=const\neq 0$  and $\bm{\pi}=\langle\bm{\pi}\rangle=0$. Using this approximation, one obtains the partition function at zero temperature and non-zero chemical potential,

\begin{align}
Z=&\displaystyle \int \mathcal{D}q\mathcal{D}\overline{q}\mathcal{D}\sigma \mathcal{D}\bm{\pi} e^{iS}\nonumber\\[0.21cm]
\vspace{1em}
=&\displaystyle \int \mathcal{D}q\mathcal{D}\overline{q}\exp\Big[i\int d^{4}x\sqrt{-g}(\overline{q}(i\gamma_{\mu}\nabla^{\mu}-M+\mu\gamma_{0})q-\frac{3}{2G}\sigma^{2})\Big]\nonumber\\[0.21cm]
\vspace{1em}
=&Z_{q}\displaystyle \exp\Big[-i\int d^{4}x\sqrt{-g}\frac{3}{2G}\sigma^{2})\Big],\label{eq:2}
\end{align}

\noindent where $ M=m+\sigma$. Since the integral of $q,\overline{q}$ is the Gaussian integral, $\log Z_{q}$ is

\begin{align}
\log Z_{q}=&\log\det[i\gamma_{\mu}\nabla^{\mu}-M+\mu\gamma_{0}]\nonumber\\[0.21cm]
\vspace{1em}
=&\log\det[(\gamma_{\mu}\nabla^{\mu})^{2}-2i\mu\nabla_{0}-\mu^{2}+M^{2}].\label{eq:zq}
\end{align}

\noindent The determinant is over spinor, color, flavor, and coordinate space. We ignored a normalization factor. (\ref{eq:zq}) in the positive curvature space, $R\times S^{3}$ (the Einstein universe) and the negative curvature space, $R\times H^{3}$ (the hyperbolic space) was calculated in Refs \cite{rf:njlg2} and \cite{rf:njlg3}.

The line element in the space-time $R\times S^{3}$ is

\begin{equation}
ds^{2}=dt^{2}-a^{2}(d\theta^{2}+\sin^{2}\theta d\Omega_{2}),
\end{equation}

\noindent where the radius of the Einstein universe $a$ is related to the scalar curvature by the relation $R=6/a^{2}$, and $d\Omega_{2}$ is the metric on the two dimensional unit sphere. (\ref{eq:zq}) in the space-time $R\times S^{3}$ is given by

\begin{equation}
\displaystyle \frac{1}{Vt}\log Z_{q}=\frac{3}{V}\int\frac{dp_{0}}{2\pi}\sum_{l=0}^{\infty}d_{l}\big(\log[(E_{l}+\mu)^{2}-p_{0}^{2}]+\log[(E_{l}-\mu)^{2}-p_{0}^{2}]\big),\label{eq:ein1}
\end{equation}

\noindent where $Vt=\displaystyle \int d^{4}x\sqrt{-g},\  l=0,1,2,\ldots,\ \xi_{l}=(l+3/2)/a$ and $E_{l}=\sqrt{\xi_{l}^{2}+M^{2}}$. The space volume $V$ and the degeneracy $d_{l}$ are expressed as

\begin{equation}
V=2\pi^{2}a^{3},\ d_{l}=2(l+2)(l+1).
\end{equation}

The line element in the space-time $R\times H^{3}$ is

\begin{equation}
ds^{2}=dt^{2}-a^{2}(d\theta^{2}+\sinh^{2}\theta d\Omega_{2}),
\end{equation}

\noindent where the radius of the hyperboloid $a$ is related to the scalar curvature by the relation $R=-6/a^{2}$. The scalar curvature in the Hyperbolic space is negative. (\ref{eq:zq}) in the space-time $R\times H^{3}$ is given by

\begin{equation}
\displaystyle \frac{1}{Vt}\log Z_{q}=6\int\frac{dp_{0}}{2\pi}\int_{0}^{\infty}d\rho\mu(\rho)(\log[(\lambda(\rho)+\mu)^{2}-p_{0}^{2}]+\log[(\lambda(\rho)-\mu)^{2}-p_{0}^{2}])\label{eq:hyper1}
\end{equation}

\noindent where $\rho=a\sqrt{\lambda^{2}-M^{2}}$ and $\mu(\rho)=(\rho^{2}+1/4)/(2\pi^{2}a^{3}).$

\vspace{1em}
\vspace{1em}
\section{$\phi$ dependent thermodynamic potential}\label{sec:phi}

To find a critical point for the chiral and deconfinement phase transitions, we consider the thermodynamic potential. The thermodynamic potential $\Omega$ is defined as

\begin{equation}
\displaystyle \Omega=-\frac{1}{\beta V}\log Z,
\end{equation}

\noindent where $\beta=1/T$ is the inverse temperature and $V$ is a space volume. However, this thermodynamic potential has no the order parameter for the deconfinement phase transition. For this reason, we calculate the dressed Polyakov loop.

We introduce the dual quark condensate. The dual quark condensate is defined as\cite{rf:dress}

\begin{equation}
\displaystyle \Sigma_{n}=-\int_{0}^{2\pi}\frac{d\phi}{2\pi}e^{-i\phi n}\langle\overline{q}q\rangle_{\phi}=\frac{3}{G}\int_{0}^{2\pi}\frac{d\phi}{2\pi}e^{-i\phi n}\sigma_{\phi}.
\vspace{1em}
\end{equation}

\noindent The $\phi$ dependence is caused by the $U(1)$ valued boundary condition $\psi(\beta,\bm{x})=e^{i\phi}\psi(0,\bm{x})$, and the Matsubara frequency, $\omega_{n}=2\pi T(n+1/2) (n=0,\pm 1,\pm 2,\ldots)$ for fermions is replaced by $\omega_{n}(\phi)=2\pi T(n+\phi/2\pi).\ \Sigma_{1}$, which is called the dressed Polyakov loop, contains the Polyakov loop. Thus, $\Sigma_{+1}$ (or $\Sigma_{-1}$) is the order parameter for deconfinement. We refer to $\sigma_{\pi}$ as to the chiral condensate for simplicity. 

The $\log$ terms in (\ref{eq:ein1}) and (\ref{eq:hyper1}) are the same as the ordinary thermodynamic potential in the NJL model. Although an imaginary part is caused by the $\phi$ dependence, we consider only a real part approximately. This derives a result similar to the lattice simulation \cite{rf:njldr}. Due to this, using the formula,

\begin{equation}
\displaystyle \sum_{n=-\infty}^{\infty}\frac{1}{x^{2}+(a+2n\pi)^{2}}=\frac{\sinh x}{2x(\cosh x-\cos a)},
\end{equation}

\noindent we can easily obtain the $\phi$ dependent thermodynamic potential for (\ref{eq:ein1}) and (\ref{eq:hyper1}). (\ref{eq:ein1}) with the $U(1)$ valued boundary condition is expressed as

\begin{align}
-\displaystyle \frac{1}{\beta V}{\rm Re}\log Z_{q}=&-\displaystyle \frac{3}{\beta V}\sum_{n=-\infty}^{\infty}\sum_{l=0}^{\infty}d_{l}(\log[\omega_{n}^{2}(\phi)+(E_{l}+\mu)^{2}]+\log[\omega_{n}^{2}(\phi)+(E_{l}-\mu)^{2}])\nonumber\\[0.21cm]
\vspace{1em}
=&-\displaystyle \frac{6}{V}\sum_{l=0}^{\infty}d_{l}(E_{l}+\sum_{\epsilon=\pm 1}\frac{T}{2}\log[1+e^{-2\beta(E_{l}+\epsilon\mu)}-2e^{-\beta(E_{l}+\epsilon\mu)}\cos\phi]),\label{eq:ein2}
\end{align}

\noindent where $E_{l}=\sqrt{\xi_{l}^{2}+M_{\phi}^{2}} ($since $M_{\phi}=m+\sigma_{\phi},\ M$ has the $\phi$ dependence). Similarly, (\ref{eq:hyper1}) with the $U(1)$ valued boundary condition is expressed as

\begin{align}
-\displaystyle \frac{1}{\beta V}{\rm Re}\log Z_{q}=&-\displaystyle \frac{6}{\beta}\sum_{n=-\infty}^{\infty}\int_{0}^{\infty}d\rho\mu(\rho)(\log[(\lambda(\rho)+\mu)^{2}+\omega_{n}^{2}(\phi)]+\log[(\lambda(\rho)-\mu)^{2}+\omega_{n}^{2}(\phi)])\nonumber\\[0.21cm]
\vspace{1em}
=&-12\displaystyle \int_{0}^{\infty}d\rho\mu(\rho)(\lambda(\rho)+\sum_{\epsilon=\pm 1}\frac{T}{2}\log[1+e^{-2\beta(\lambda+\epsilon\mu)}-2e^{-\beta(\lambda+\epsilon\mu)}\cos\phi]).
\end{align}

\noindent Instead of $\rho$, using the dimension of momentum $p=\rho/a$ (we will omit the ${\rm Re}$) \cite{rf:njlg3},

\begin{equation}
-\displaystyle \frac{1}{\beta V}\log Z_{q}=-\frac{6}{\pi^{2}}\int_{0}^{\infty}dp(p^{2}+\frac{1}{4a^{2}})(E_{p}+\sum_{\epsilon=\pm 1}\frac{T}{2}\log[1+e^{-2\beta(E_{p}+\epsilon\mu)}-2e^{-\beta(E_{p}+\epsilon\mu)}\cos\phi]),\label{eq:hyper2}
\end{equation}

\noindent where $E_{p}=\sqrt{p^{2}+M_{\phi}^{2}}.$

The real parts of $\phi$ dependent thermodynamic potentials for the space-time $R\times S^{3}$ and $R\times H^{3}$ are written as

\begin{align}
&\displaystyle \Omega_{\phi}^{ein}=\frac{3}{2G}\sigma_{\phi}^{2}-\frac{6}{V}\sum_{l=0}^{\infty}d_{l}(E_{l}+\sum_{\epsilon=\pm 1}\frac{T}{2}\log[1+e^{-2\beta(E_{l}+\epsilon\mu)}-2e^{-\beta(E_{l}+\epsilon\mu)}\cos\phi]),\label{eq:ein3}\\[0.28cm]
\vspace{1em}
&\displaystyle \Omega_{\phi}^{hyp}=\frac{3}{2G}\sigma_{\phi}^{2}-\frac{6}{\pi^{2}}\int_{0}^{\infty}dp(p^{2}+\frac{1}{4a^{2}})(E_{p}+\sum_{\epsilon=\pm 1}\frac{T}{2}\log[1+e^{-2\beta(E_{p}+\epsilon\mu)}-2e^{-\beta(E_{p}+\epsilon\mu)}\cos\phi]).\label{eq:hyper3}
\end{align}

\vspace{1em}
\vspace{1em}
\section{Numerical calculation}\label{sec:nu}

Since (\ref{eq:ein3}) and (\ref{eq:hyper3}) are divergent at large $l$ and $p$, we should introduce a cutoff. We introduce the multiplier $e^{-\omega_{l}/\Lambda}$  as the cutoff in summation over $l$ in (\ref{eq:ein3}) (due to the present lack of experimental knowledge, the cutoff in the space-time $R\times S^{3}$ can not be determined), and the momentum cutoff $\Lambda$ in (\ref{eq:hyper3}). Furthermore, to make dimensional quantities dimensionless, we denote the notation $\Omega/\Lambda^{4}\rightarrow\Omega, \sigma/\Lambda\rightarrow\sigma,  T/\Lambda\rightarrow T,\cdots$. Then, the regularized thermodynamic potentials are written as

\begin{align}
\displaystyle \Omega_{\phi}^{ein}=\frac{3}{2G}\sigma_{\phi}^{2}-\frac{6}{V}\sum_{l=0}^{\infty}e^{-\omega_{l}}d_{l}(E_{l}+\sum_{\epsilon=\pm 1}\frac{T}{2}\log[1+e^{-2\beta(E_{l}+\epsilon\mu)}-2e^{-\beta(E_{l}+\epsilon\mu)}\cos\phi]),\label{eq:ein4}\\[0.28cm]
\vspace{1em}
\displaystyle \Omega_{\phi}^{hyp}=\frac{3}{2G}\sigma_{\phi}^{2}-\frac{6}{\pi^{2}}\int_{0}^{1}dp(p^{2}+\frac{1}{4a^{2}})(E_{p}+\sum_{\epsilon=\pm 1}\frac{T}{2}\log[1+e^{-2\beta(E_{p}+\epsilon\mu)}-2e^{-\beta(E_{p}+\epsilon\mu)}\cos\phi]).\label{eq:hyper4}
\end{align}

\noindent$G,\ R,\ T,\ \mu$, and $\sigma_{\phi}$ in this chapter and figures represent dimensionless parameters $G\Lambda^{2},\ R/\Lambda^{2},\  T/\Lambda,\ \mu$/$\Lambda$, and $\sigma_{\phi}/\Lambda$. In the chiral limit, one can obtain a critical point to calculate the global minimum of these. When $\Omega(M_{\pi}\neq 0)$ has a minimum value, chiral symmetry is broken, and when $\Omega(M_{\pi}=0)$ has a minimum value, chiral symmetry is restored.

The gap equations are given by

\begin{equation}
\displaystyle \frac{\partial\Omega_{\phi}}{\partial\sigma_{\phi}}=0.
\end{equation}

\noindent In the chiral limit, when $M_{\pi}=\sigma_{\pi}=0$ is obtained from this equation, chiral symmetry is restored. Due to this, ordinary chiral condensate $\sigma_{\pi}$ is the order parameter for the chiral symmetry. However, since the chiral symmetry is broken explicitly in the case of $m\neq 0$, i.e., the critical temperature is defined by the peak of the chiral susceptibility,

\begin{equation}
\displaystyle \chi=\frac{\partial\sigma_{\pi}}{\partial T}.
\end{equation}

\noindent Similarly, using the dressed Polyakov loop, a critical temperature for the deconfinement phase transition is defined by

\begin{equation}
\displaystyle \tau=\frac{\partial\Sigma_{1}}{\partial T}.
\end{equation}

\begin{figure}[t]

\begin{center}

\includegraphics[width=75mm]{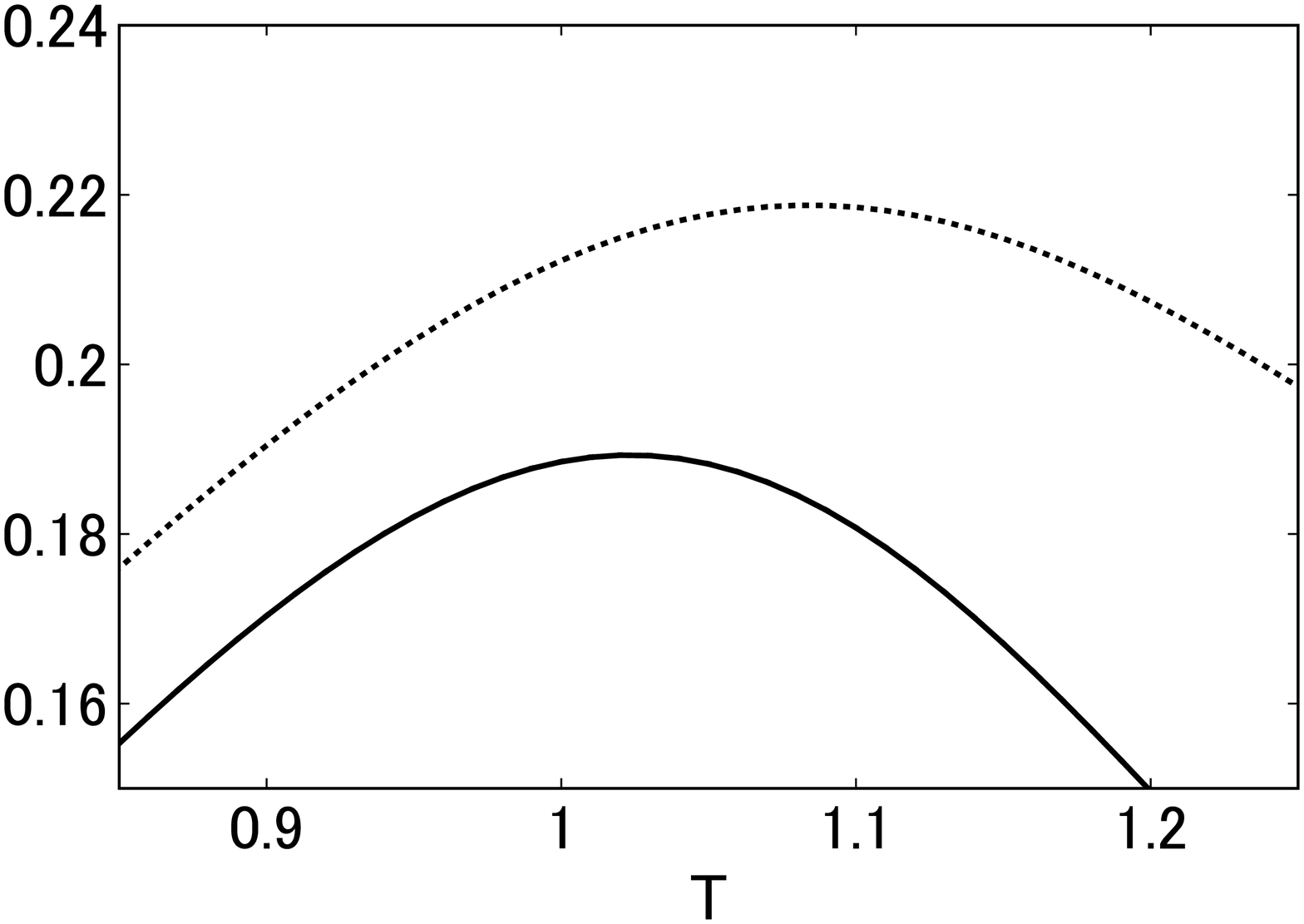}

\caption{$\chi$ and $\tau$ at around peaks. The solid line indicates $-0.1\chi$, and the dotted line indicates $\tau.\ G=10,\ m=0.1,\ R=10$, and $\mu=0.5$}

\label{fig:peak}

\end{center}

\end{figure}

\noindent For decision on a critical temperature, we considered order $10^{-3}$ for a temperature. However, for example, as shown in Fig.\ \ref{fig:peak}, the critical temperature has an error range from the determination of peak. (When $m=0$, the peak width is very narrow.) Taking these into account, the critical temperature should have a numerical error of order $10^{-2}\sim 10^{-3}$ at least.

We use a large value for the dimensionless current mass $m$. If $u$ and $d$ quarks are used in the flat space-time, the dimensionless current mass is very small (see section \ref{sec:a} or \cite{rf:hatsu,rf:Buba}). However, when setting a small dimensionless current mass, the difference between the chiral and the deconfinement phase transitions is small. Due to this, we use $m=0.1$ for the space-time $R\times S^{3} $and $R\times H^{3}$.

The summation and integration in (\ref{eq:ein4}) and (\ref{eq:hyper4}) converge at curvature $R<1$ in our calculation. The multiplier $ e^{-\omega_{l}}/\Lambda$ is small enough and the momentum cutoff is larger than the gravitational scale for $R<15$. In addition, the chiral phase transition in the chiral limit is observed at high curvature, and $\sigma_{\phi}$ and $\Sigma_{1}$ connect from low curvature ($R<1$) to high curvature $(R>1)$ smoothly.

\vspace{1em}
\subsection{The positive curvature space-time $R\times S^{3}$}

In the flat space-time, the chiral phase transition is of the second order (the crossover) in the high temperature region in the case of $m=0$ ($m\neq 0$), and is of the first order in the low temperature region in cases of $m=0$ and $m\neq 0$. The deconfinement phase transition is the crossover in the high temperature region, and is of the first order in the low temperature region \cite{rf:njldr}.

\begin{figure}[t]

\begin{center}

\includegraphics[width=75mm]{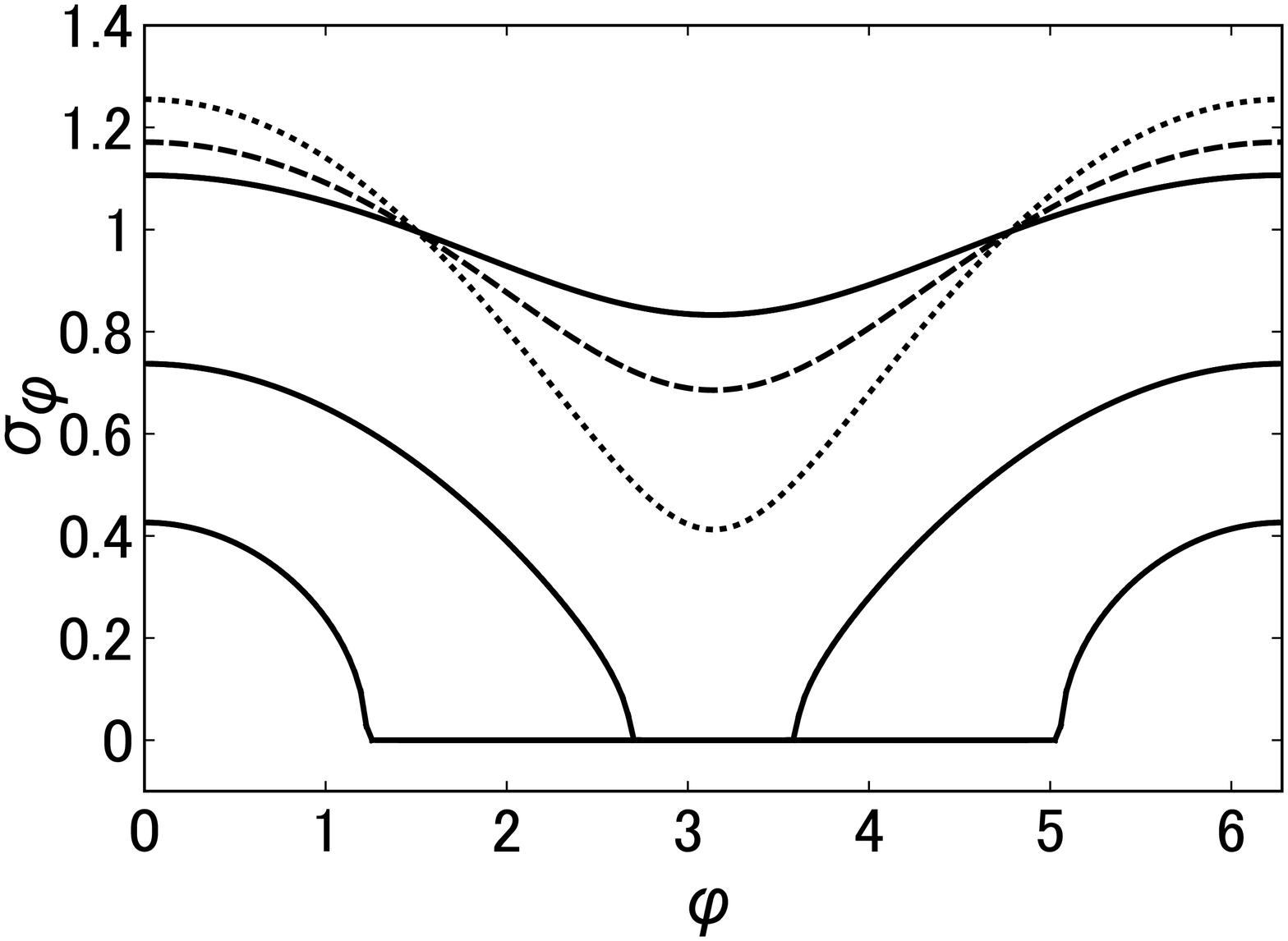}

\caption{$\phi$ dependence of the $\sigma_{\phi}$ in the case of $m=0$. From the top to the bottom the solid lines correspond to $R=12,\ R=13.3$, and $R=14$ at $T=0.5$ and $\mu=0.5$, respectively. The dashed line and the dotted line correspond to $T=0.65, \mu=0.5$ and $T=0.5, \mu=0.8$ at $R=12$, respectively.}

\label{fig:Mphi1}

\end{center}

\end{figure}

\begin{figure}[t]

\begin{center}

\includegraphics[width=75mm]{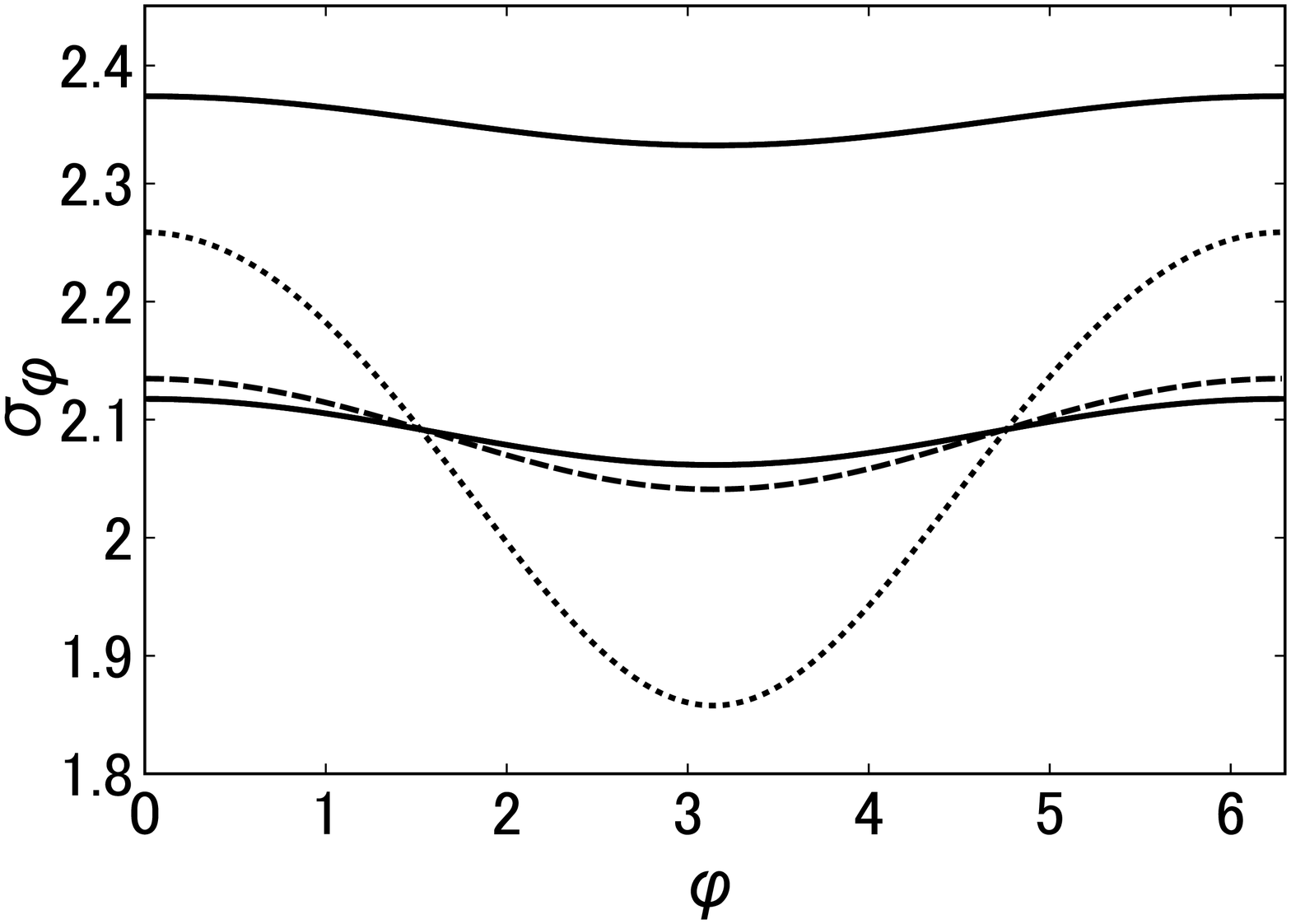}

\caption{$\phi$ dependence of the $\sigma_{\phi}$ in the case of $m=0.1$. From the top to the bottom the solid lines correspond to $R=5$ and $R=7$ at $T=0.5$ and $\mu=0.5$, respectively. The dashed line and the dotted line correspond to $T=0.8, \mu=0.5$ and $T=0.5, \mu=0.8$ at $R=7$, respectively.}

\label{fig:Mphi2}

\end{center}

\end{figure}

\begin{figure}[t]

\begin{center}

\includegraphics[width=75mm]{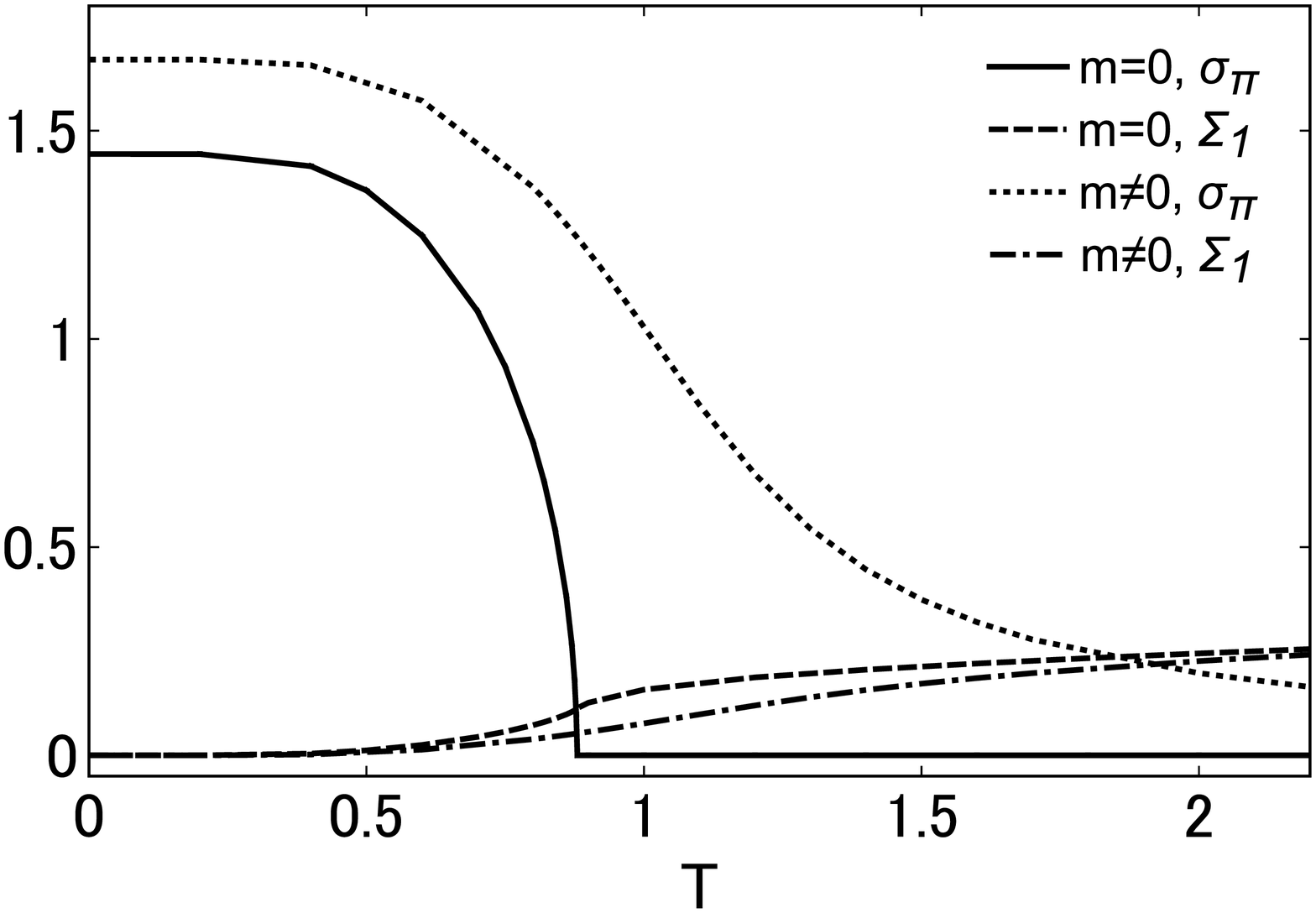}

\caption{The chiral condensate $\sigma_{\pi}$ and the dressed Polyakov loop $\Sigma_{1}$ in cases of $m=0$ and $m=0.1$ at $R=10$ and $\mu=0.5$.}

\label{fig:M-dress1}

\end{center}

\end{figure}

Similarly, we calculate the $\phi$ dependent chiral condensate $\sigma_{\phi}$ and the dressed Polyakov loop $\Sigma_{1}$ in the space-time $R\times S^{3}$. The value of the parameter $G$ is taken as $10$. We perform the numerical calculation for two cases, $m=0$ and $m\neq 0$. Figs.\ \ref{fig:Mphi1} and \ref{fig:Mphi2} show the $\phi$ dependence of the $\sigma_{\phi}$ for $m=0$ and $m=0.1$, respectively. $\sigma_{\phi}$ decreases with increasing curvature $R$. When compared to the change in temperature, the curvature simply lowers the value of $\sigma_{\phi}$. The chiral condensate $\sigma_{\pi}$ and the dressed Polyakov loop $\Sigma_{1}$ in cases of $m=0$ and $m=0.1$ are shown Fig.\ \ref{fig:M-dress1}. Fig.\ \ref{fig:M-dress1} shows that the chiral phase transition in the case of $m=0$ ($m\neq 0$) is of the second order (crossover), and the deconfinement phase transition is the crossover in case of $m=0$ and $m\neq 0$. On the other hand, since the chiral condensate and the dressed polyakov loop have discontinuous values at the critical points in the low temperature and high chamical potential region in the same way as the flat space-time, these phase transitions are of the first order.

\begin{figure}[t]

\begin{center}

\includegraphics[width=75mm]{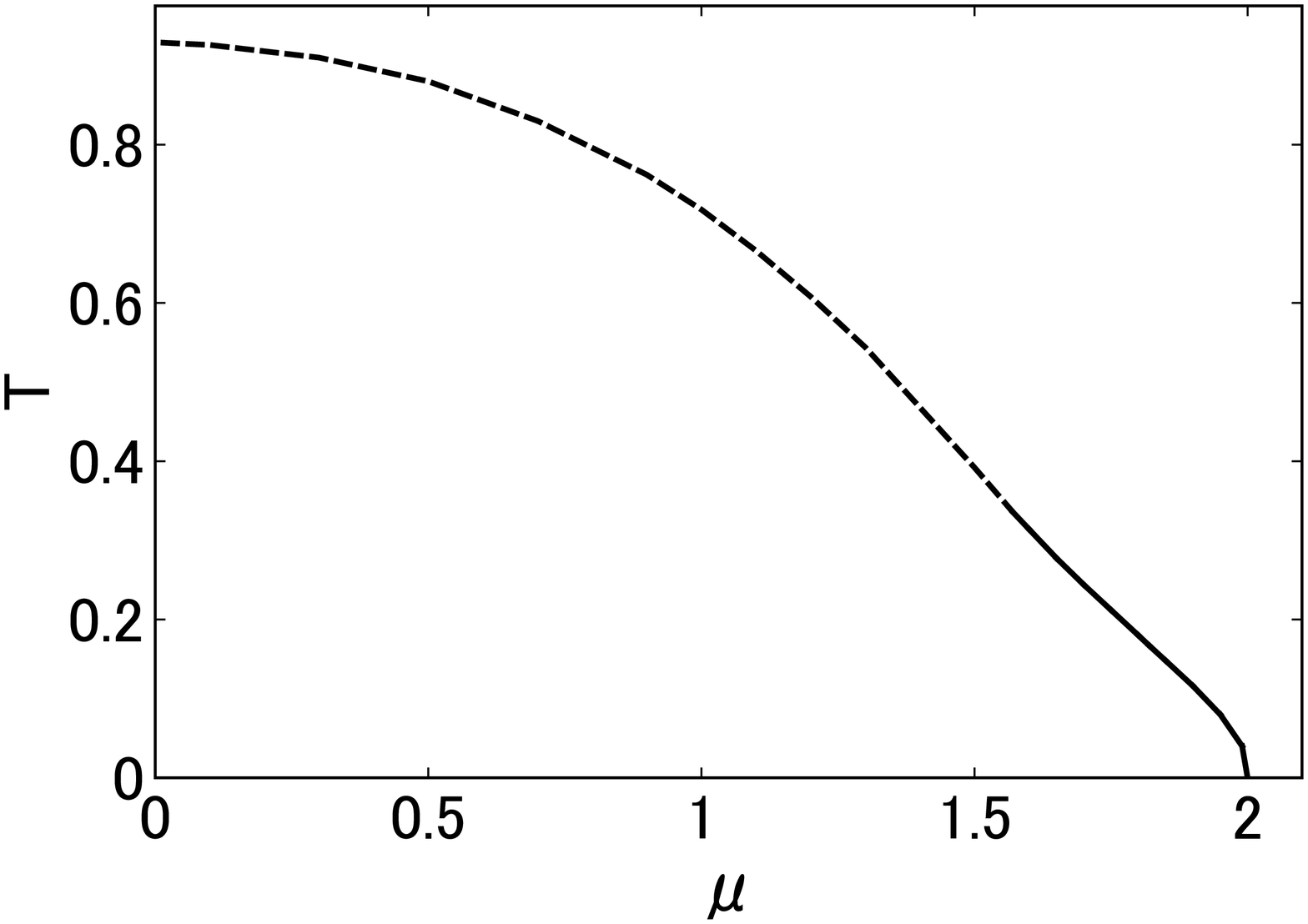}

\caption{$ T-\mu$ phase diagram for the case of $m=0$. The solid line indicates the first order phase transition, and the dashed line indicates the second order phase transition at $R=10$.}

\label{fig:Tmu1}

\end{center}

\end{figure}

\begin{figure}[t]

\begin{center}

\includegraphics[width=75mm]{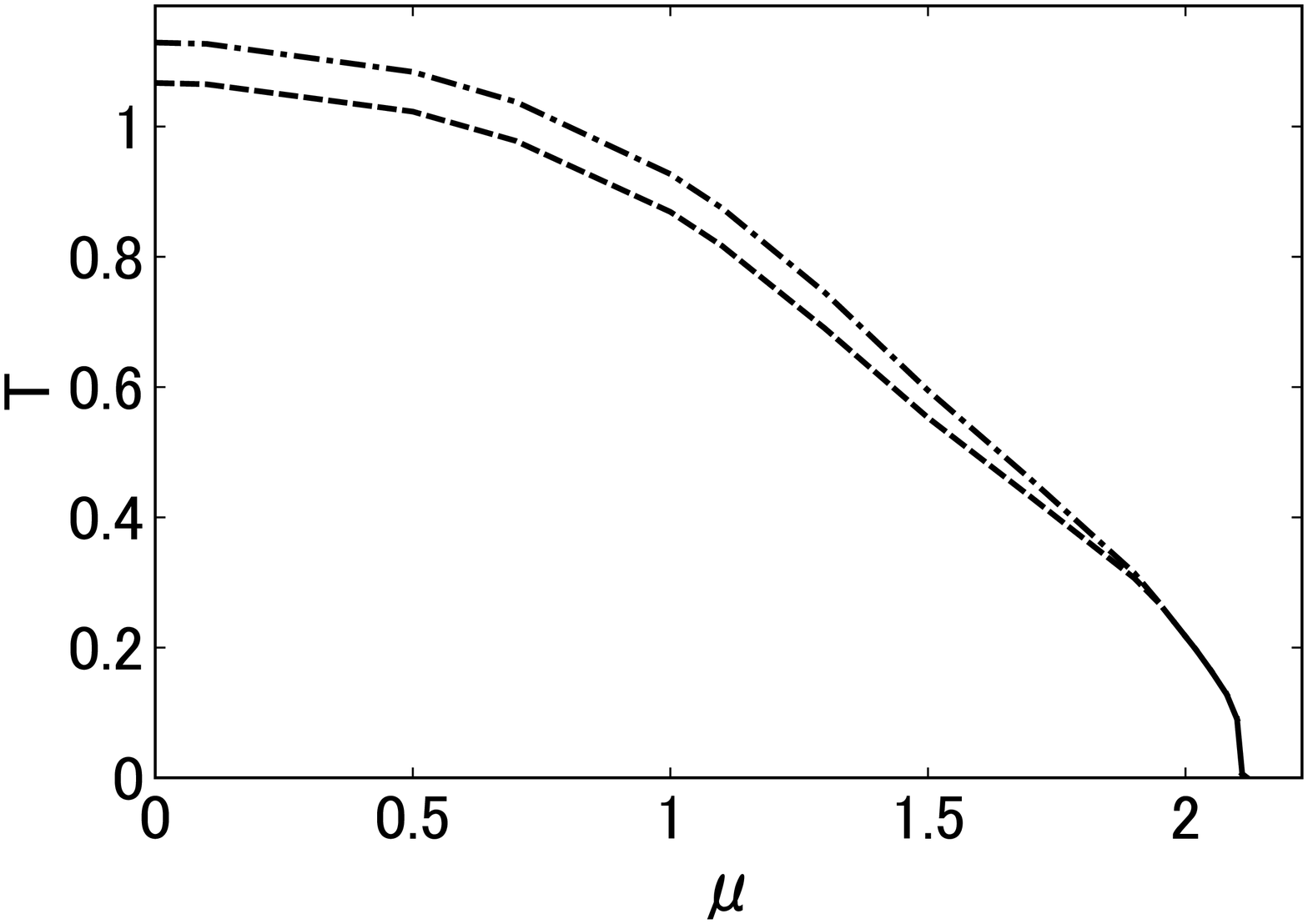}

\caption{$ T-\mu$ phase diagram for the case of $m=0.1$. The solid line indicates the first order phase transition, and the dashed and dash-dotted line indicate the crossover at $R=10$. The dashed line indicates the chiral condensate, and the dash-dotted line indicates the dressed polyakov loop.}

\label{fig:Tmu2}

\end{center}

\end{figure}

\begin{figure}[t]

\begin{center}

\includegraphics[width=75mm]{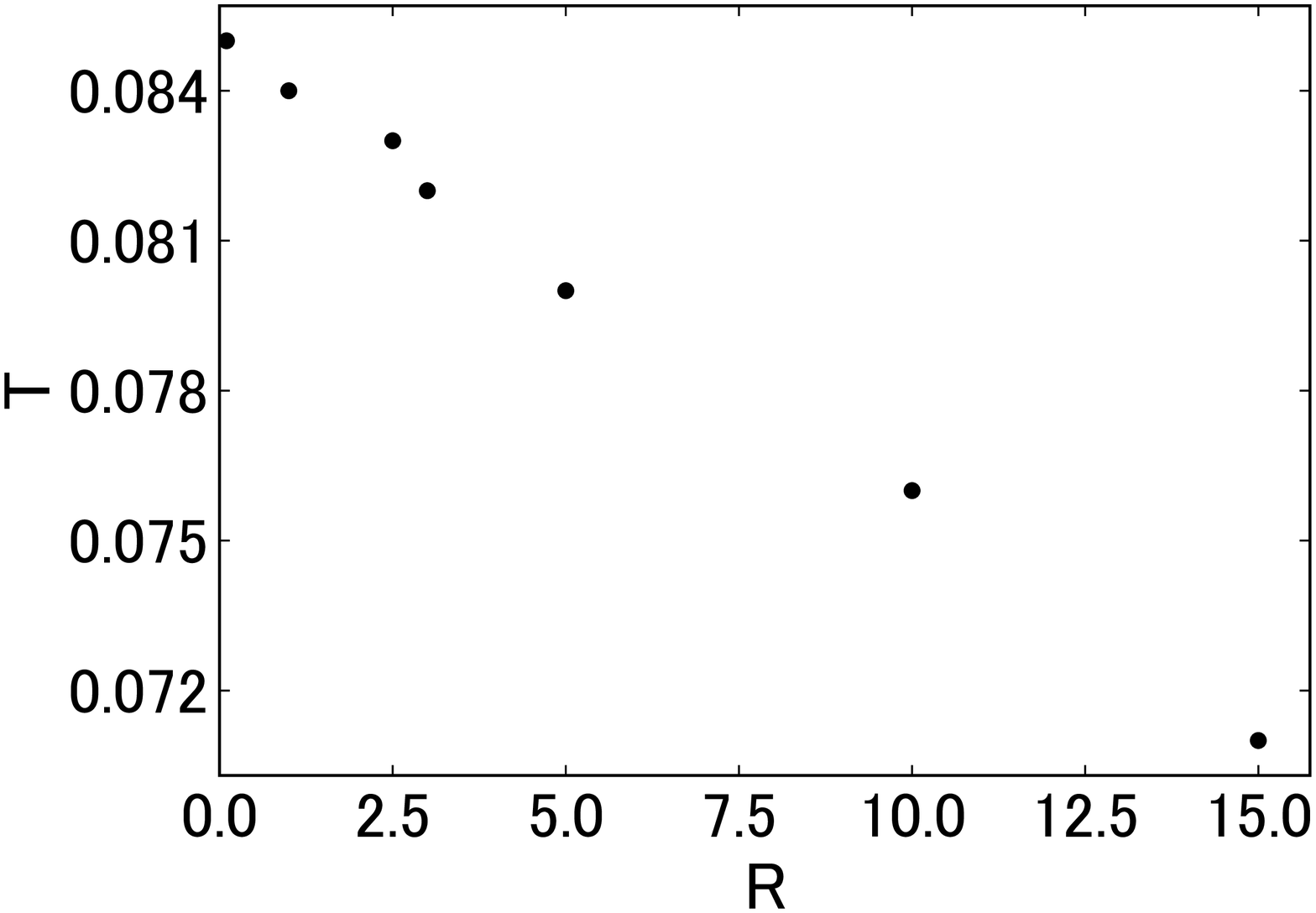}

\caption{Circle marks indicate the difference $T_{c}^{d}-T_{c}^{\chi}$ . $G=15,\ m=0.1$, and $\mu=0$.}

\label{fig:gap1}

\end{center}

\end{figure}

The critical points for the chiral phase transition are identical with the deconfinement phase transition in the case of $m=0$, even there is the curvature (Fig.\ \ref{fig:Tmu1}). By contrary, Fig.\ \ref{fig:Tmu2} shows that the critical points for the chiral phase transition are not identical with the deconfinement phase transition in the case of $m=0.1$. To clarify the difference, $G$ is taken as $15$ in Figs.\ \ref{fig:Tmu2} and \ref{fig:gap1}. The NJL model in the ordinary four dimensional flat space-time includes this property. However, the difference gets smaller by increasing curvature. In addition, when $m=0.4$, the difference is $0.14$ at $R=0.1$, and is $0.115$ at $R=15$. Thus, the difference gets larger by increasing the current mass. On the other hand, since the change of the difference by increasing curvature is very small and the peak widths of $\chi$ and $\tau$ are not narrow compared to the change (see Fig.\ \ref{fig:peak}), our result bears uncertainty. However, the change of the difference by increasing curvature is identified within the range of our numerical error. For this reason, the difference should change by increasing curvature.

\begin{figure}[t]

\begin{center}

\includegraphics[width=75mm]{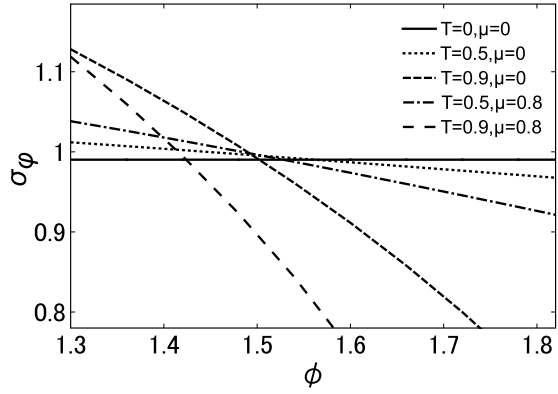}

\caption{$\sigma_{\phi}$ ($m=0$) at around $\phi=\pi/2$. The point at $\phi=\pi/2$ is not invariant under changes in $T$ and $\mu$.}

\label{fig:no}

\end{center}

\end{figure}

Note that the points at around $\phi=\pi/2$ in Figs.\ \ref{fig:Mphi1} and \ref{fig:Mphi2} are not invariant under changes in $T$ and $\mu$. Actually, these points depend to $T$ and $\mu$ weakly (see Fig.\ \ref{fig:no}). The contribution terms in $\sigma_{\phi}$ for $\phi=0,$\ $\pi/2,$\ $\pi$ are given by

\begin{align*}
\displaystyle \phi=0\ &:\displaystyle \ 1-\frac{1}{e^{\beta(E-\mu)}-1}-\frac{1}{e^{\beta(E+\mu)}-1},\\[0.21cm]
\vspace{1em}
\displaystyle \phi=\pi/2\ &:\displaystyle \ 1-\frac{1}{e^{2\beta(E-\mu)}+1}-\frac{1}{e^{2\beta(E+\mu)}+1},\\[0.21cm]
\vspace{1em}
\displaystyle \phi=\pi\ &:\displaystyle \ 1-\frac{1}{e^{\beta(E-\mu)}+1}-\frac{1}{e^{\beta(E+\mu)}+1}.
\end{align*}

\noindent Thus, due to the factor $2$, when compared to $\sigma_{0}$ and $\sigma_{\pi},\ \sigma_{\pi/2}$ may not change significantly under changes in $T$ and $\mu$.

\vspace{1em}
\vspace{1em}
\subsection{The negative curvature space-time $R\times H^{3}$}\label{sec:a}

\begin{figure}[t]

\begin{center}

\includegraphics[width=75mm]{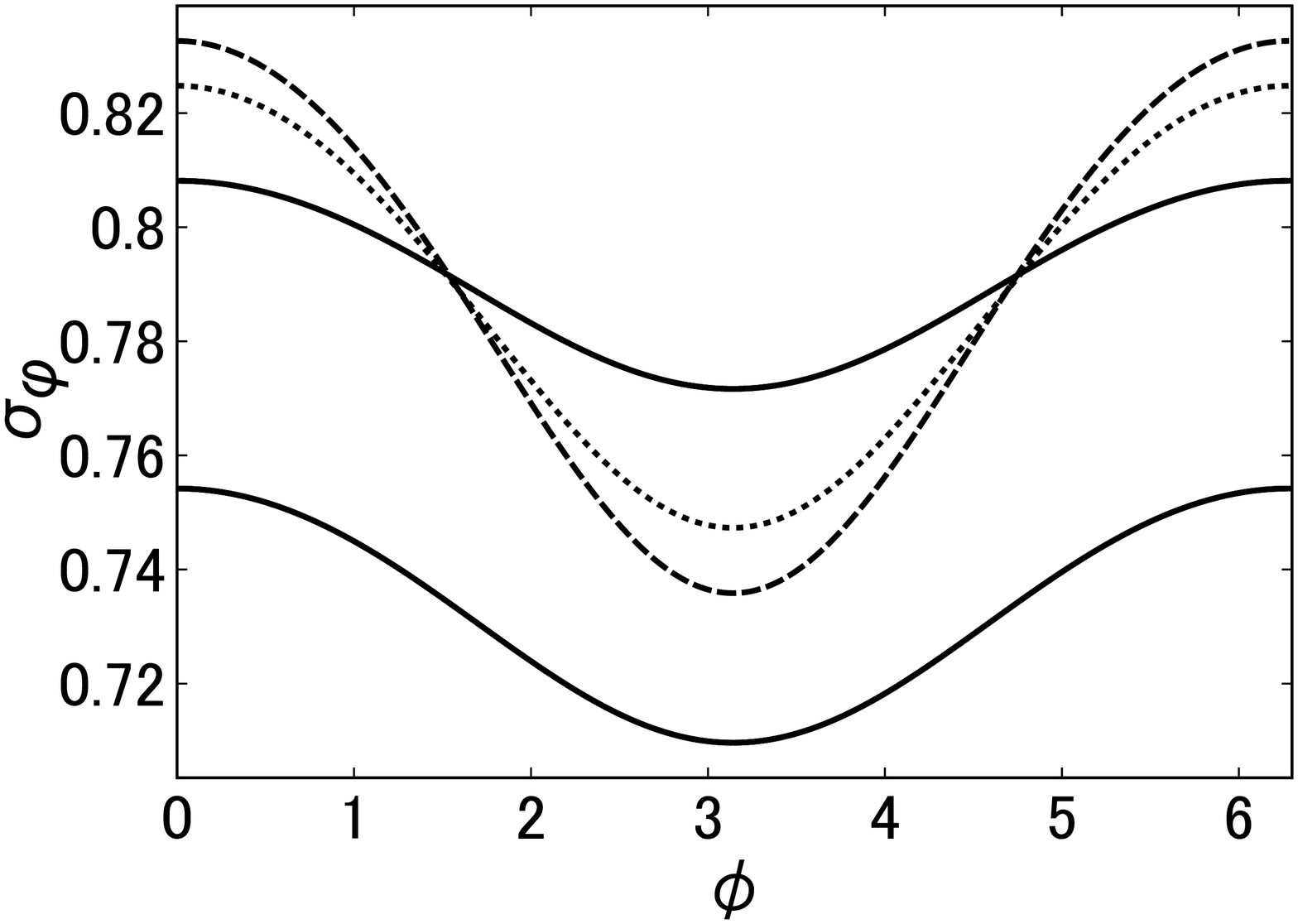}

\caption{$\phi$ dependence of the $\sigma_{\phi}$ in the case of $m=0$. From the top to the bottom the solid lines correspond to $|R|=0.5$ and $|R|=0.2$ at $T=0.2$ and $\mu=0.1$, respectively ($R<0$). The dashed line and the dotted line correspond to $T=0.25,\ \mu=0.1$, and $T=0.2,\ \mu=0.3$ at $|R|=0.5$, respectively.}

\label{fig:Mphi-nega1}

\end{center}

\end{figure}

\begin{figure}[t]

\begin{center}

\includegraphics[width=75mm]{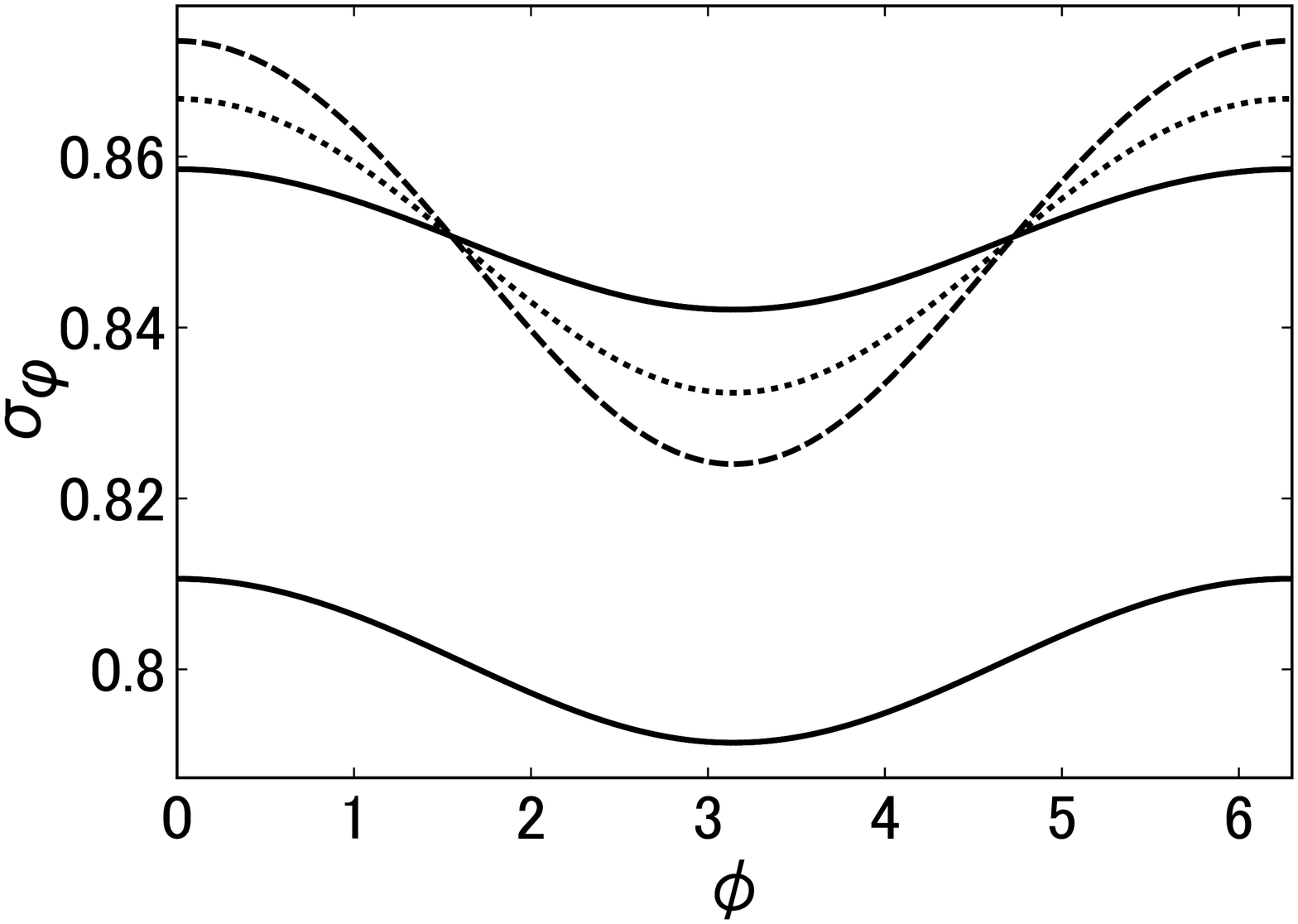}

\caption{$\phi$ dependence of the $\sigma_{\phi}$ in the case of $m=0.1$. From the top to the bottom the solid lines correspond to $|R|=0.2$ and $|R|=0.5$ at $T=0.2$ and $\mu=0.1$, respectively ($R<0$). The dashed line and the dotted line correspond to $T=0.25,\ \mu=0.1$ and $T=0.2,\ \mu=0.3$ at $|R|=0.5$, respectively.}

\label{fig:Mphi-nega2}

\end{center}

\end{figure}

\begin{figure}[t]

\begin{center}

\includegraphics[width=75mm]{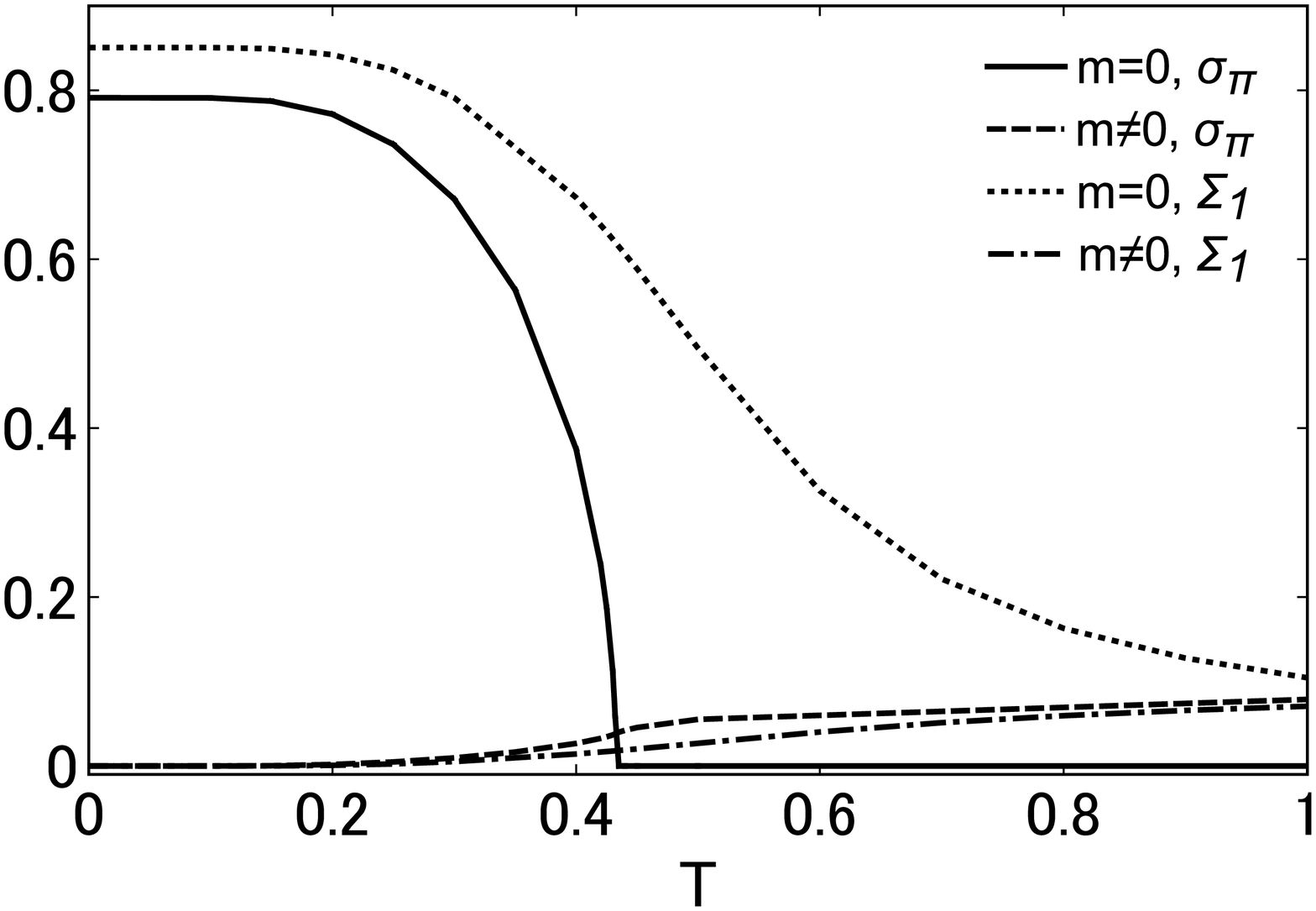}

\caption{The chiral condensate $\sigma_{\pi}$ and the dressed Polyakov loop $\Sigma_{1}$ in cases of $m=0$ and $m=0.1$ at $R=-0.5$ and $\mu=0.1$.}

\label{fig:M-dress2}

\end{center}

\end{figure}

Chiral symmetry is broken for an arbitrary small coupling constant in the space-time $R\times H^{3}$, and the chiral condensate depends nonanalytical on the curvature \cite{rf:njlg3}. However, we consider only the case of strong coupling. $G$ is taken as $15$. Figs.\ \ref{fig:Mphi-nega1} and \ref{fig:Mphi-nega2} show the $\phi$ dependence of the $\sigma_{\phi}$ for $m=0$ and $m=0.1$, respectively. Since $\sigma_{\phi}$ increases with increasing curvature $|R|$, the negative curvature yields the opposite effect. The chiral condensate $\sigma_{\pi}$ and the dressed Polyakov loop $\Sigma_{1}$ in cases of $m=0$ and $m=0.1$ are shown Fig.\ \ref{fig:M-dress2}. 

\begin{figure}[t]

\begin{center}

\includegraphics[width=75mm]{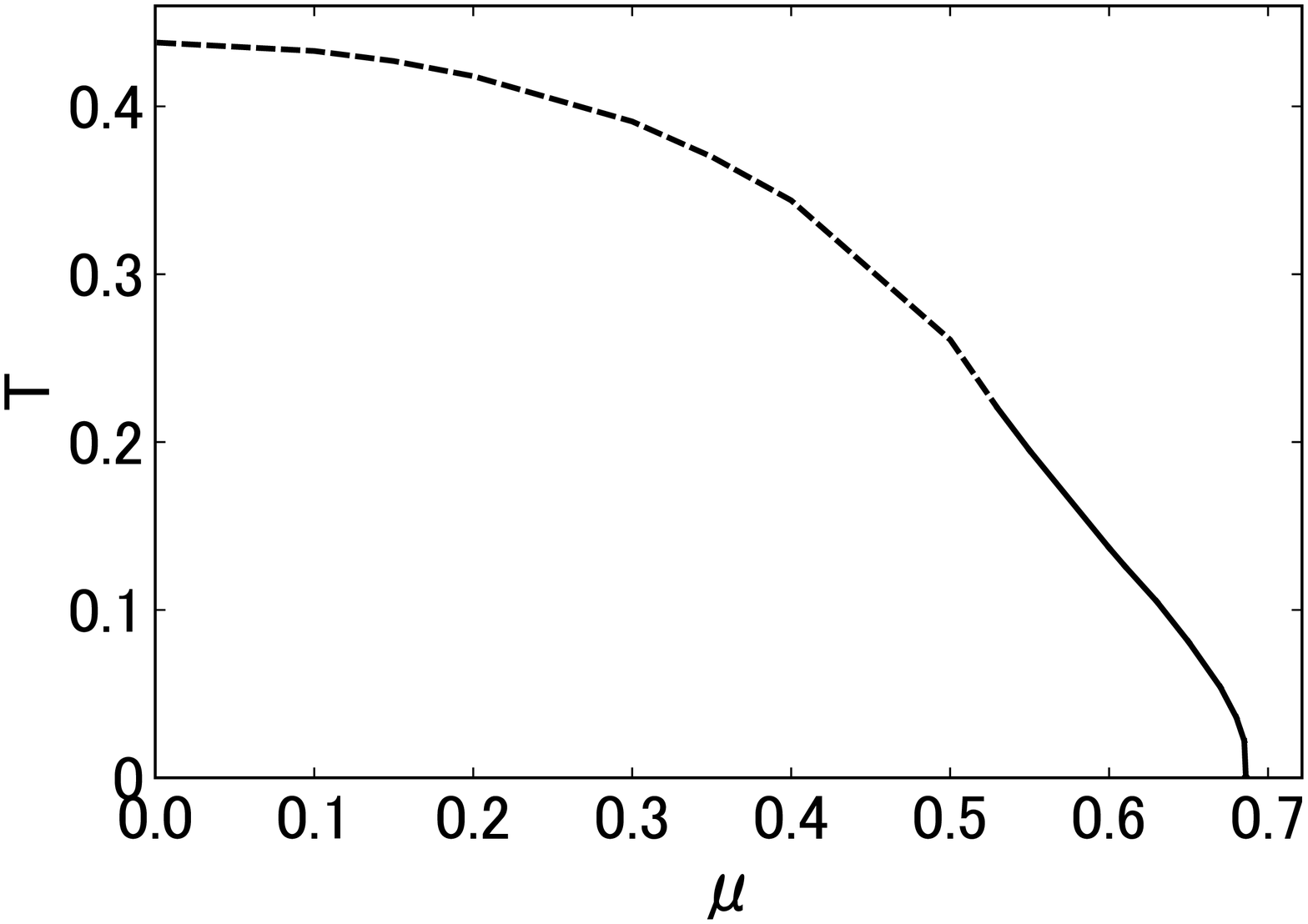}

\caption{$ T-\mu$ phase diagram for the case of $m=0$. The solid line indicates the first order phase transition, and the dashed line indicates the second order phase transition at $R=-0.5$.}

\label{fig:Tmu-nega1}

\end{center}

\end{figure}

\begin{figure}[t]

\begin{center}

\includegraphics[width=75mm]{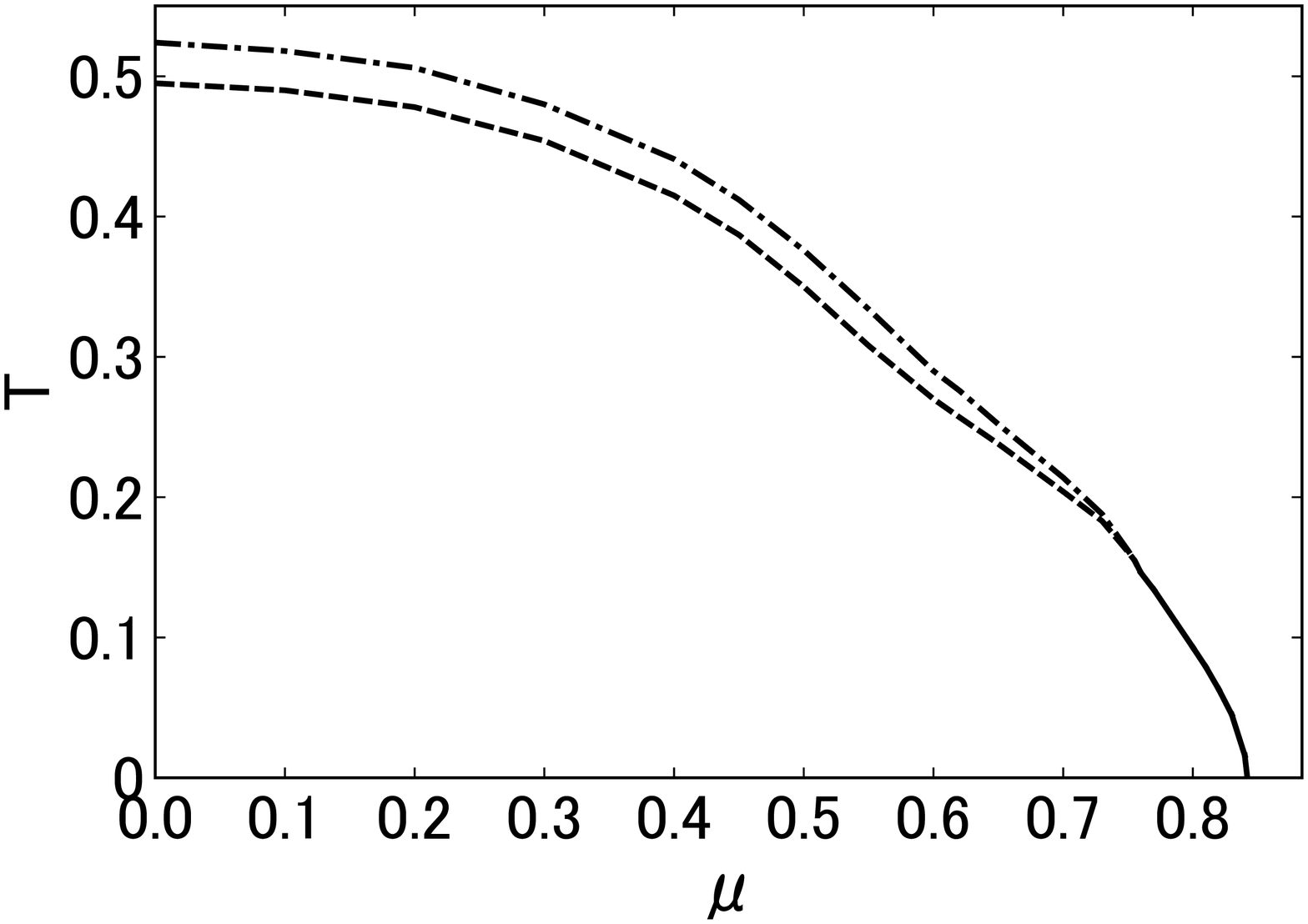}

\caption{$ T-\mu$ phase diagram for the case of $m=0.1$. The solid line indicates the first order phase transition, and the dashed and dash-dotted lines indicate the crossover at $R=-0.5$. The dashed line indicates the chiral condensate, and the dash-dotted line indicates the dressed polyakov loop.}

\label{fig:Tmu-nega2}

\end{center}

\end{figure}

\begin{figure}[t]

\begin{center}

\includegraphics[width=75mm]{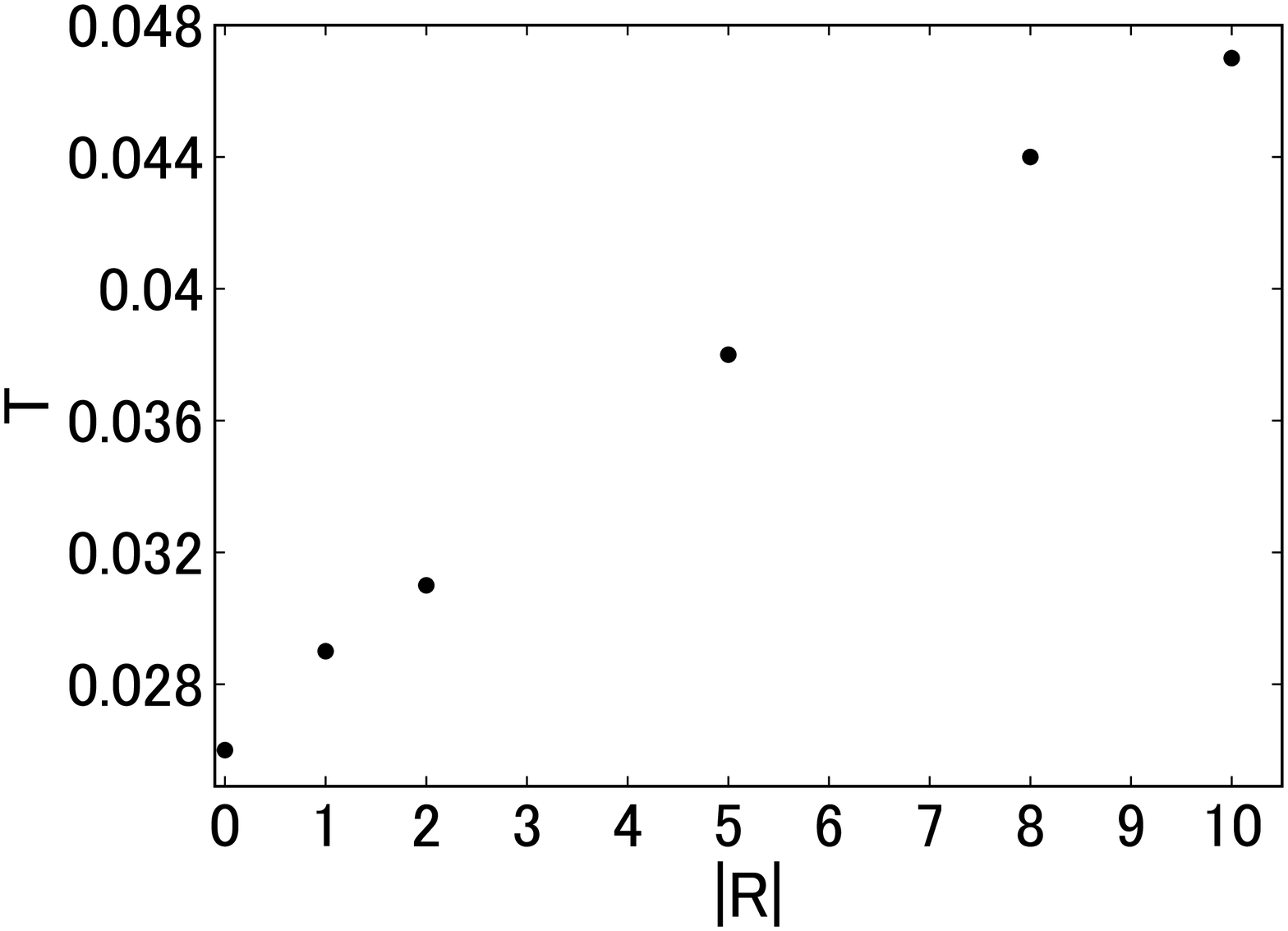}

\caption{Circle marks indicate the difference $T_{c}^{d}-T_{c}^{\chi}.\ G=15,\ m=0.1$, and $\mu=0$.}

\label{fig:gap-nega}

\end{center}

\end{figure}

Also the negative curvature does not make the difference between the critical points in the case of $m=0$ (Fig.\ \ref{fig:Tmu-nega1}). By contrary, in the case of $m=0.1$, the negative curvature increases the difference (Figs.\ \ref{fig:Tmu-nega2} and \ref{fig:gap-nega}).

Unlike the space-time $R\times S^{3}$, since the thermodynamic potential (\ref{eq:hyper4}) in the space-time $R\times H^{3}$ corresponds to one in the flat space-time, one should be able to use the cutoff $\Lambda$ in the flat space-time. The cutoff $\Lambda$ in the flat space-time is estimated in the usual manner \cite{rf:Buba}. For example, using the pion mass $135$ MeV and pion decay constant $92.4$ MeV, we obtain $\sigma_{\pi}\Lambda=392$ MeV with  $G=14.6,\ m\Lambda=5.6$ MeV, and $\Lambda=588$ MeV in the flat space-time. However, since we used the large value for the current mass, the current mass does not correspond to these, and is not realistic. Thus, we show values for the purpose of reference. Using $G=15$ and $m=0.1$, since we obtain $\Lambda=317.3$ MeV in the flat space-time, at $\mu=0$, the difference of temperature in the space-time $R\times H^{3}$ is $8.2$ MeV at $|R|=0,\ 9.2$ MeV at $|R|=1$, and $14.9$ MeV at $|R|=10$.

\vspace{1em}
\vspace{1em}
\section{Summary}\label{sec:sum}

In this paper, using the NJL model in the curved space, we investigated whether the chiral and deconfinement phase transitions are separated by the gravitational effect. Then, we used the dressed Polyakov loop as the order parameter for deconfinement. We tried the positive curvature space-time $R\times S^{3}$, and the negative curvature space-time $R\times H^{3}$. In both cases, the critical points are not different in the chiral limit $m=0$. By contrast, the critical points are different for the case of $m\neq 0$ in the crossover region. The NJL model in the flat space has this property. However, the difference decreases (increases) with increasing curvature in the space-time $R\times S^{3}$ ($R\times H^{3}$). Thus, the gravity should have a different effect on the ordinary chiral condensate and the dressed polyakov loop in the crossover region.

Thus, the difference for the critical points in the flat space is caused by the presence or absence of the current mass $m$, and the gravity induces changes of the difference. In addition, a mass should relate to the gravity. For this reason, the ordinary chiral condensate and the dressed polyakov loop have a different dependence for the current mass in the crossover region (the difference increases with the increasing current mass in the flat space), and the positive (negative) curvature reduces (enhances) the contribution of the current mass. Incidentally, since the difference with increasing curvature for small values of $m$ is very small, light particles are scarcely affected by the gravitational effect. Thus, the gravitational effect should become effective for heavy particles, such as $s$ quark.

Due to using the model having no a gluon, only a quark is taken into account in this paper. However, QCD has the gluon dynamics, and a gluon relates to confinement. Thus, to get more understanding of the gravitational effect for the chiral and deconfinement phase transitions, we must investigate QCD. In particular, the gluon may be affected by the gravity.

\vspace{1em}
\vspace{1em}
\section*{Acknowledgements}

This work was partially supported by the Research Center for Measurement in Advanced Science of Rikkyo University.

\newpage

\vspace{1em}
\vspace{1em}
\vspace{1em}

\vspace{1em}

\vspace{1em}

\vspace{1em}

\vspace{1em}
\vspace{1em}

\vspace{1em}

\vspace{1em}
\vspace{1em}

\vspace{1em}
\vspace{1em}

\vspace{1em}
\vspace{1em}

\vspace{1em}

\vspace{1em}

\vspace{1em}

\vspace{1em}

\vspace{1em}
\vspace{1em}

\end{document}